\documentclass[12pt,a4paper]{article}

\usepackage{latexsym}
\usepackage{amsmath}
\usepackage{amsfonts}
\usepackage{amssymb}

\newcommand{\be}{\begin{equation}}
\newcommand{\ee}{\end{equation}}
\newcommand{\bea}{\begin{eqnarray}}
\newcommand{\eea}{\end{eqnarray}}

\def\tr{{\mathrm{tr}}}           
\def\jp{\frac{1}{2}}             
\def\js{\frac{1}{4}}             
\def\bC{{\mathbb C}}             
\def\gln{GL(n,\mathbb{C})}       
\def\1{{\mbox{\boldmath $1$}}}    %
\def\bR{{\mathbb R}}             
\def\bT{{\mathbb T}}             
\def\C{{\mathcal C}}             
\def\ri{{\mathrm i}}             
\def\cn{{\mathcal N}}            

\begin{document}

\vspace*{0.5cm}
\begin{center}
{\Large \bf Poisson-Lie generalization of the
Kazhdan-Kostant-Sternberg reduction}

\end{center}

\vspace{0.2cm}

\begin{center}
L. Feh\'er${}^{a,}$
and C. Klim\v c\'\i k${}^b$ \\

\bigskip

${}^a$Department of Theoretical Physics, MTA  KFKI RMKI\\
1525 Budapest, P.O.B. 49,  Hungary, and\\
Department of Theoretical Physics, University of Szeged\\
Tisza Lajos krt 84-86, H-6720 Szeged, Hungary\\
e-mail: lfeher@rmki.kfki.hu

\medskip

${}^b$Institute de math\'ematiques de Luminy,
 \\ 163, Avenue de Luminy, \\ 13288 Marseille, France\\
 e-mail: klimcik@iml.univ-mrs.fr

\bigskip

\end{center}

\vspace{0.2cm}

\begin{abstract} The trigonometric Ruijsenaars-Schneider model  is derived
by symplectic reduction of  Poisson-Lie symmetric   free motion on the  group  $U(n)$.
 The
 commuting flows of the model are effortlessly obtained by reducing  canonical free flows on the Heisenberg
double of $U(n)$.
The free flows are associated with a very simple Lax matrix, which is shown to yield
the Ruijsenaars-Schneider Lax matrix upon  reduction.

\bigskip


\medskip


\end{abstract}

\newpage

\section{Introduction}
\setcounter{equation}{0}
\sloppy \raggedbottom

The    integrable many-body models of Calogero-Moser-Sutherland type
\cite{Cal,Sut,Mos} enjoy huge popularity in mathematical physics, for they
possess a wide range of physical applications based on their
intimate relationships to central areas of harmonic analysis, theory
of special functions
 and symplectic geometry
 (see \cite{SR-CRM} for a review).
 One of the fruitful approaches to study the  structure of these models
 is to represent them
 in terms of Hamiltonian reductions   of `free'  systems  with large  symmetries.
 This philosophy  originated from the celebrated papers of  Olshanetsky-Perelomov \cite{OP1,OP2} and
 Kazhdan-Kostant-Sternberg \cite{KKS}
 who, among others,   derived the trigonometric  Sutherland model  by reduction
 of the free particle moving on the group $U(n)$.

\medskip

\noindent  In this paper, we shall develop a Poisson-Lie generalization of the
Kazhdan-Kostant-Sternberg reduction and we shall show that it yields
the trigonometric  Ruijsenaars-Schneider
relativistic integrable system  \cite{RS}\footnote{To be more precise, we shall deal with
the standard one  out of the possible physically different real forms  \cite{SR-RIMS} of the
complex trigonometric Ruijsenaars-Schneider model.}.
 Thus  our result
 corroborates the general expectation  that  the reduction
 translates the transition from   the
  ordinary to the Poisson-Lie  symmetric free systems    into the transition from
  the non-relativistic to the relativistic
 integrable many-body models.

\medskip

\noindent
We here recall that various infinite dimensional generalizations  of the Kazhdan-Kostant-Sternberg reduction
have been studied  in the nineties by
Gorsky-Nekrasov and others (\cite{GN,Nekr,Fock+} and references therein).
In particular, in the paper \cite{GN}  a
trigonometric Ruijsenaars-Schneider model
was derived by reducing a Hamiltonian system on
the magnetic cotangent bundle of the loop group of $U(n)$.
This reduction is intrinsically close to topological field theory and thus it should have
a finite dimensional counterpart.
In fact,
Gorsky and Nekrasov \cite{GN}  stated it as a problem to work out such a finite dimensional reduction
in terms of a Heisenberg double.
This motivated us, although we shall derive the usual trigonometric
Ruijsenaars-Schneider model with Hamiltonian (\ref{RS-Ham}), which is
different from the $\mathrm{III}_{\mathrm{b}}$ model \cite{SR-RIMS, DV} model obtained in \cite{GN}.

  \medskip

  \noindent Arguably, the crucial point in the Kazhdan-Kostant-Sternberg
  derivation of the trigonometric Sutherland  model was
  the choice  of a certain  element $\iota(x)$ of the Lie algebra $u(n)$  of the group
  $U(n)$, defined
  by
  \be \iota(x)_{jj}=0,\quad \forall j,\qquad \iota(x)_{jk}=\ri x, \quad \forall j\neq k,
  \label{jx}\ee
  where $x$ is a real non-zero parameter.
  Kazhdan, Kostant and Sternberg  took for the unreduced phase space
  just the cotangent bundle $T^*U(n)$,  they picked as  the unreduced Hamiltonian the one
  which induces  the Killing geodesics on $U(n)$, and they reduced using the adjoint action of
  $U(n)$ on $T^*U(n)$ by constraining the moment map $J$ of this  action to be equal to $\iota(x)$:
  \be
  J(K)=\iota(x), \quad K\in T^*U(n).
  \label{KKS}\ee
  One may view
the element $\iota(x)$ as the key needed to unlock the room
  inside  $T^*U(n)$ in which the Sutherland model is stored, since  starting from the
  constraint (\ref{KKS})   the standard symplectic reduction procedure  yields the
  trigonometric Sutherland model \cite{KKS}.

  \medskip

  \noindent
The Poisson-Lie analogue of the cotangent bundle  $T^*U(n)$ is  the
well-known Heisenberg double of $U(n)$ constructed by
Semenov-Tian-Shansky in 1985 \cite{ST}. What is somewhat less known
is the correct Poisson-Lie analogue of the adjoint action of $U(n)$
on $T^*U(n)$, but this was made explicit by one of us in a recent
paper \cite{K}.
The Heisenberg double is naturally equipped with
a Hamiltonian that is invariant with respect to
the  `quasi-adjoint' action  of  \cite{K} and generalizes
the kinetic energy of the free geodesic motion.
Thus the  true problem is to identify  the  Poisson-Lie analogue of
the Kazhdan-Kostant-Sternberg element $\iota(x)$  which will permit to find
the trigonometric Ruijsenaars-Schneider model inside the Heisenberg double.
Our claim is that this analogue is
  the upper-triangular $n\times n$ matrix $\nu(x)$ given by
\be \nu(x)_{jj}=1, \quad \forall j,\qquad  \nu(x)_{jk}= (1-e^{-x})e^{\frac{(k-j)x}{2}},
\quad \forall j<k.\label{bx}\ee
The quasi-adjoint action has a so-called Poisson-Lie moment map which is a map $\Lambda$
from the Heisenberg double $D$ to the group $B$ of complex  upper-triangular $n\times n$ matrices with
positive real numbers on the diagonal. The Poisson-Lie reduction is then   determined
by requiring
\be
\Lambda(K)=\nu(x),\quad K\in D.
\label{PLKKS'}\ee
As in the standard Kazhdan-Kostant-Sternberg case, the reduction is completely algorithmic,
although it is technically more involved to solve the constraint (\ref{PLKKS'}) and
to identify the reduced system.
After doing this below, we shall also explain how we have found the constant $\nu(x)$.
Readers not familiar with symplectic reduction based on Poisson-Lie symmetries
may study  Lu's work \cite{Lu-red}, but to understand our paper it really
suffices to know that
the reduction tool operates  in the same way as for ordinary symmetries if
a `good moment map' is available.

\medskip

\noindent   We organize  the   present paper as follows.
In Section 2, we describe the Poisson-Lie symmetric  free motion
 on the group $U(n)$. Then, in  Section 3, we obtain the trigonometric Ruijsenaars-Schneider
 model and its Lax matrix by
 reducing   the free flows on the Heisenberg double.
The reduction crucially uses the statement of Theorem 1, which is  subsequently proved
in a technically involved
  Section 4.
  Finally, in Section 5, we further compare the approach of this paper
  with the earlier related results in the literature
  and also
   briefly touch on
  some aspects of the Ruijsenaars duality
that will be described in detail elsewhere.

 \section {Poisson-Lie symmetric free motion on the group $U(n)$}
 \setcounter{equation}{0}

  The Heisenberg double of $U(n)$  is  the group $\gln$ viewed as a {\it real} manifold.
  Every point $K\in \gln$ admits two unambiguous Iwasawa decompositions
 \be
K= b_L g_R^{-1}
\quad\hbox{and}\quad
K= g_L b_R^{-1}
\quad\hbox{with}\quad
b_{L,R}\in B,\,\, g_{L,R}\in U(n).
\label{N.7}\ee
   There is   a
 natural symplectic form $\omega_+$  on $\gln$, first described in \cite{AM},
  \be
  \omega_{+}=\jp \Im \tr(d\Lambda_L\Lambda_L^{-1}\wedge d\Xi_L\Xi_L^{-1})+
  \jp\Im \tr(d\Lambda_R\Lambda_R^{-1}\wedge d\Xi_R\Xi_R^{-1}).
  \label{ST}\ee
Here $\Im z$ stands for the imaginary part of the complex number $z$, $\tr$ is  the ordinary
matrix trace  and we use  the  Iwasawa maps
$\Lambda_{L,R}: \gln \to B$ and
$\Xi_{L,R}: \gln \to U(n)$ given by
\be
\Lambda_{L,R}(K):= b_{L,R}
\quad\hbox{and}\quad
\Xi_{L,R}(K):= g_{L,R}.
\label{Iwasawa}\ee

\medskip

 \noindent
 Define now the Hermitian matrix-valued function $L$ on $\gln$  as
 \be
 L(K):=(K^\dagger K)^{-1} = \Lambda_R(K) \Lambda_R(K)^\dagger,
 \quad K\in\gln.
 \label{Lorig}\ee
 By  Poisson-Lie  symmetric free motion on  $U(n)$ we mean a dynamical system the phase space of which is the
 Heisenberg double of $U(n)$ and the Hamiltonian  of which is provided by
 \be
 H_\mu(K):= \jp\sum_{j\neq 0}\frac{\mu_j}{j}\tr(L(K)^{j}), \quad K\in\gln.
 \label{origH}\ee
 Whatever is the sequence  of the real parameters $\mu_j$, the Hamiltonian $H_\mu$ is
 obviously invariant with respect to the
following quasi-adjoint  action $\triangleright $ introduced in \cite{K}:
  \be
  g  \triangleright K:=gK\Xi_R(g\Lambda_L(K)), \qquad g\in U(n),\quad  K\in \gln.
  \label{QAD}\ee
The flow induced by this Hamiltonian was explicitly identified in  \cite{Zakr}  and it reads
\be
K_\mu (t)=b\exp{\biggl(-\ri t \sum_{j\neq 0} \mu_j(b^\dagger b)^{-j}\biggr)}g^{-1}.
\label{flows}\ee
Here  the constant elements $b\in B$ and $g\in U(n)$   encode  the choice of initial conditions.
If we interpret  $b_L$ in the decomposition (\ref{N.7}) as `momentum' and $g_R$
as  `position', then (\ref{flows}) says
that the momentum is conserved and the position follows the standard Killing
geodesics on $U(n)$. This fact justifies the terminology `free motion'.

\medskip

\noindent If  $\mu$ and $\mu'$ are two arbitrary sequences of real parameters then
it follows easily from the  formula (\ref{flows})  that the corresponding flows commute.
In other words,
we have
\be
\{H_\mu,H_{\mu'}\}_+=0,
\label{comm}\ee
where  $\{.,.\}_+$ is the Poisson bracket induced by the
symplectic form $\omega_+$.
Hence  $L$ (\ref{Lorig}) can be interpreted
as the Lax matrix of the commuting family of the dynamical systems
 $(\gln, \omega_+,H_\mu)$.

\section{The reduction}
\setcounter{equation}{0}
In \cite{K}, the quasi-adjoint action  $\triangleright$ was   shown to admit
the so-called Poisson-Lie moment map
 $\Lambda:\gln\to B$ given by
 \be
 \Lambda(K)=\Lambda_L(K)\Lambda_R(K), \quad K\in \gln.
 \label{product} \ee
 This means, in particular, that for every $X\in u(n)$, every $K\in \gln$ and every
 function $f$ on $\gln$, we have
 \be
 {d\over ds}f(e^{sX} \triangleright  K)\vert_{s=0}=\Im\tr(X\{f,\Lambda\}_+(K)\Lambda(K)^{-1}).
 \ee

 \medskip

 \noindent The first step of the reduction of the Poisson-Lie symmetric free
  motion $(\gln,\omega_+,H_\mu)$  amounts
 to solving the moment map  constraint
 \be
 \Lambda(K)=\Lambda_L(K)\Lambda_R(K)=\nu(x),\quad K\in \gln,
 \label{PLKKS}\ee
with $\nu(x)$ defined in (\ref{bx}).
 The result is summarized in the following theorem.

 \medskip

 \noindent {\bf Theorem 1.}  \emph{Denote by $\C$ the set of the regular elements  of a Weyl alcove
 in the maximal torus $\bT_n\subset U(n)$,
  by $A$  the diagonal subgroup  of $B$, by $N$ the group of  complex upper-triangular   matrices
  having $1$ all along the diagonal, and by $G_x$
  the isotropy group of $\nu(x)$, i.e.,
\be
  G_x:=\{g\in U(n)\,\vert\,g\nu(x)\nu(x)^\dagger g^{-1}=\nu(x)\nu(x)^\dagger \}.
  \label{Gx}\ee
  Then  every solution $K$  of the moment map constraint (\ref{PLKKS})
  can be written as
\be
K=g \triangleright (\cn(T)aT^{-1}),
\label{csol}\ee
where     $T\in \C$, $a\in A$, $g\in G_x$  and
   $\cn(T)\in N$  is given by
 \be
 \cn(T)_{kl}=\prod_{m=1}^{l-k}
\frac{e^{\frac{x}{2}}T_l-e^{-\frac{x}{2}}
 T_{k+m}}{T_l- T_{k+m-1}},
\qquad \forall k<l.
\label{4.21}\ee
Moreover, it holds that no two different points  of
the form $\cn(T)aT^{-1}$ can be transformed into each other by the action of $G_x$.
}

 \medskip

 \noindent  The message of Theorem 1 is that the submanifold of $\gln$ defined as
 \be
 S:=\{\cn(T)aT^{-1}\,\vert\, T\in \C, a\in A \}
 \label{Slice}\ee
 forms a global cross section of the orbits of $G_x$ in the inverse image of the moment map value $\nu(x)$.
 Therefore  $S$ can serve as a model
 of the resulting reduced phase space.

 \medskip

\noindent  In order not to break  the line of
 the presentation, we postpone the  proof of Theorem 1 for a while and, as the second step of the reduction,
we evaluate the symplectic form (\ref{ST}) on the slice $S$ (\ref{Slice}).
 The calculation  of the
Iwasawa maps (\ref{Iwasawa}) on the slice  gives directly
\bea
\Lambda_L(\cn(T)aT^{-1})=\cn(T)a,\quad \Xi_R(\cn(T)aT^{-1})=T, \\
\Lambda_R(\cn(T)aT^{-1})= T a^{-1}\cn(T)^{-1}T^{-1},\quad
\Xi_L(\cn(T)aT^{-1})=T^{-1},
\eea
and, consequently, the reduced symplectic form reads
\be
\omega_r(T,a)=\Im\tr(T^{-1}dT\wedge a^{-1}da) .
\ee
We choose the following parametrization of  $T$ and $a$:
\be
T:={\rm diag}(e^{2\ri q_1},e^{2\ri q_2},...,e^{2\ri q_n}),
\qquad 0\leq q_k < \pi, \quad q_1> q_2> ...> q_n;
\label{strict}\ee
\be
a:={\rm diag}(e^{\zeta_1},e^{\zeta_2},...,e^{\zeta_n}),
\ee
where
\be
\zeta_k= -\frac{p_k}{2}-\js \sum_{m<k}{\rm ln}\biggl(1 +
\frac{\sinh^2\frac{x}{2}} { \sin^2(q_k - q_m)}  \biggr)+\js \sum_{m>k}{\rm ln}\biggl(1 +
\frac{\sinh^2\frac{x}{2}} { \sin^2(q_k - q_m)}  \biggr).
\ee
Then the reduced symplectic form $\omega_r$ becomes the Darboux one
\be
\omega_r =\sum_{k} dp_k \wedge dq_k.
\ee
Hence we can identify the reduced phase space with the cotangent bundle $T^*{\cal C}$
of the open Weyl alcove $\cal C$.

\medskip

\noindent On account of (\ref{origH}),
the  reduced Hamiltonians are given by the formula
\be
H_\mu(T,a)=\jp \sum_{j\neq 0}\frac{\mu_j}{j}\tr(L(T,a)^j),
\label{rham}\ee
where
\be
 L(T,a)=a^{-1}\cn(T)^{-1}({\cn(T)^\dagger})^{-1} a^{-1}.
 \label{Lax}\ee
Since the commutativity (\ref{comm}) of the Hamiltonians is preserved by the reduction procedure, we may interpret
$L(T,a)$ as the Lax matrix of the reduced system. The  components of the Lax matrix $L$ in the
Darboux variables can be directly evaluated from (\ref{4.21}) and (\ref{Lax}). This gives
\be
L_{kl}=\frac{\Gamma_k\bar\Gamma_l e^{\frac{p_k+p_l}{2}}{\rm sinh}\frac{x}{2}}{{\rm sinh}(\frac{x}{2}+\ri q_k-\ri q_l)}
\prod_{m\neq k}\biggl(1 + \frac{\sinh^2\frac{x}{2}} { \sin^2(q_k - q_m)}  \biggr)^{1\over 4}\prod_{s\neq l}\biggl(1 +
\frac{\sinh^2\frac{x}{2}} { \sin^2(q_l - q_s)}  \biggr)^{1\over 4},
\label{xxx}\ee
where the  $U(1)$ -valued quantities $\Gamma_k$ are defined as the phase factors of the following complex numbers:
\be
 e^{-\ri q_k}\prod_{m>k}\frac{e^{-\frac{x}{2}}e^{-2\ri q_k}-e^{\frac{x}{2}}e^{-2\ri q_m}}{
e^{-2\ri q_k}-e^{-2\ri q_m}}.
\ee
In the calculation of the Lax matrix, we have used that the inverse of the matrix $\cn(T)$ is
 \be
 ( \cn(T)^{-1})_{kl}=\prod_{m=1}^{l-k}
\frac{e^{-\frac{x}{2}}\bar T_k-e^{\frac{x}{2}}
 \bar T_{k+m-1}}{\bar T_k- \bar T_{k+m}},
\qquad \forall k<l.
\label{4.22}
\ee
The   role of the Lax matrix is to generate the
commuting Hamiltonians by its eigenvalues.  Thus  the matrix obtained by the conjugation $L\to
\Gamma^{-1}L\Gamma:= \mathbf{L}$,
where $\Gamma$ is the diagonal matrix with the components $\Gamma_k$,
serves equally well as a  Lax matrix of the reduced system. We have
\be
\mathbf{L}_{kl}=\frac{e^{\frac{p_k+p_l}{2}}{\rm sinh}\frac{x}{2}}{{\rm sinh}(\frac{x}{2}+\ri q_k-\ri q_l)}
\prod_{m\neq k}\biggl(1 +
\frac{\sinh^2\frac{x}{2}} { \sin^2(q_k - q_m)}  \biggr)^{1\over 4}\prod_{s\neq l}\biggl(1 +
\frac{\sinh^2\frac{x}{2}} { \sin^2(q_l - q_s)}  \biggr)^{1\over 4},
\label{xxy}\ee
which is nothing but the  standard trigonometric Ruijsenaars-Schneider  Lax matrix \cite{RS, SR-RIMS, SR-CRM}.

\medskip

\noindent
The above arguments prove that our Hamiltonian reduction yields
 precisely the trigonometric Ruijsenaars-Schneider model.
In particular,
if we pick  $\mu_{\pm 1}=\pm 1$ and all other  $\mu_j$
vanishing, then we obtain  from (\ref{rham})  the  trigonometric  Ruijsenaars-Schneider
Hamiltonian \cite{RS, SR-RIMS, SR-CRM},
\be
H_{\mathrm{trigo-RS}}(q, p)= \sum_{k=1}^n (\cosh p_k) \prod_{m\neq k}\biggl(1 +
\frac{\sinh^2\frac{x}{2}} { \sin^2(q_k - q_m)}  \biggr)^{1\over 2}.
\label{RS-Ham}\ee
Let us finally evaluate the flows of the reduced Hamiltonians (\ref{rham}).
For any curve $g_R(t)$ in the set of the  regular elements in $G$, denote
by ${\cal E}[g_R(t)]$ its  diagonalized form varying in the Weyl alcove
$\C\subset \bT_n$.  Suppose that the
 flow (\ref{flows}) starts on the slice $S$ (\ref{Slice}) at time $t=0$. We then
  immediately obtain that the Ruijsenaars-Schneider variable $T\in \C$
develops according to
\be
T(t) =e^{2\ri q(t)}=
{\cal E}[T(0) e^{\ri t \sum_{j\neq 0} \mu_j ({{\mathbf L}}(0))^{j}}],
\label{4.41}\ee
i.e.,  $e^{2\ri q(t)}$ moves along the ordered
eigenvalues of the geodesic $e^{2\ri q(0)} e^{ \ri t \sum_{j\neq 0} \mu_j ({\mathbf{L}}(0))^{j}} $ defined
by the initial values of the coordinates and the Lax matrix ({\ref{xxy}).
This reproduces the well-known interpretation of the
commuting Ruijsenaars-Schneider flows \cite{RS, SR-CMP, SR-RIMS}.

\medskip

\noindent
It is worth stressing that the Ruijsenaars-Schneider coupling constant arises
from the arbitrary parameter $x$ in the moment map value $\nu(x)$ (\ref{bx}).
 In (\ref{RS-Ham}) the velocity of light has been set to $1$, and actually one can  vary it arbitrarily
  by performing the reduction using an arbitrary multiple of the
symplectic form $\omega_+$ (\ref{ST}) on the Heisenberg double.

\section{The proof of Theorem 1}
\setcounter{equation}{0}
The proof of Theorem 1   will be based on the following three lemmas:

 \medskip

 \noindent {\bf Lemma 1.}  \emph{If a vector $v\in \bR^n$ with non-negative components,
 a real non-zero number $x\in \bR$ and an element $\nu$ of the group $N$ satisfy
 the relation
 \be
 {\rm ln}(\nu\nu^\dagger)=x(v  v^\dagger-\1_n),
 \label{rel}\ee
 then it holds
 \be
 \nu_{jk}=(1-e^{-x})e^{(k-j)x\over 2}, \quad \forall j<k\ee
 and
 \be v_k=\sqrt{n(e^x-1)\over 1-e^{-nx}}e^{-{kx\over 2}},\quad \forall k.
 \label{vx}\ee}

  \medskip

 \noindent {\bf Proof.}   First of all,   by exponentiating   Eq. (\ref{rel}),  we obtain

 \be
 \nu\nu^\dagger
=e^{-x}\left[ \1_n+ \frac{e^{nx}-1}{n}v  v^\dagger \right].
\label{3.13}\ee
Here we use the fact that $v^\dagger v=n$, which is an immediate consequence of the assumptions of the lemma.

\medskip

\noindent
Second, for an  $n\times n$ matrix $M$, denote by $M_k$
the $(n-k)\times (n-k)$ matrix
obtained by  deleting the first $k$ rows and  the first $k$ columns of
$M$ (in particular $M_0=M$).  We find
\be
\det (\nu\nu^\dagger)_{k-1}=e^{(k-1-n)x}
\left[1+\frac{e^{nx}-1}{n}\sum_{m=k}^n \vert v_m\vert^2\right]
\qquad \forall k=1,\ldots,n.
\label{3.18}\ee
We have calculated these determinants by using
  the identity
\be
\det(\1_m + u   w^\dagger) = 1 + w^\dagger u,
\qquad
\forall u, w\in \bC^m,
\label{3.17}\ee
where $u, w$  are column  vectors and
$\1_m$ is the $m\times m$  unit matrix.
\medskip

\noindent
By comparing $\det (\nu\nu^\dagger)_k$ and $\det(\nu\nu^\dagger)_{k-1}$, we can evaluate the absolute
values of all components of the vector $v$:
\be
\vert v_k\vert^2=\frac{n}{e^{nx}-1}e^{(n-k)x}\left[e^x
\det(\nu\nu^\dagger)_{k-1}-\det(\nu\nu^\dagger)_k\right], \quad  k=1,\ldots,n-1.
 \label{3.19}\ee
The upper-triangularity of any $\nu\in N$ implies
$(\nu\nu^\dagger)_{k-1}=\nu_{k-1}\nu^\dagger_{k-1}$
for $k=1,\ldots,n$,
 while the property
$\nu_{kk}=1$  implies
$\det(\nu_{k-1})=\det(\nu^\dagger_{k-1})=1$.
Therefore  $\det(\nu\nu^\dagger)_{k-1}=1$ for each $k=1,\ldots,n$,
and, from (\ref{3.19}), we conclude
\be
\vert v_k\vert^2=\frac{n(e^x -1)}{1-e^{-nx}}e^{-kx}
\qquad \forall k=1,\ldots,n.
\ee
So far we have proved the uniqueness of the non-negative real vector $v$
given by (4.3). It is
easy to check that $\nu$ given by (4.2) verifies (4.1), or, equivalently,
(4.4). The uniqueness of $\nu\in N$ then follows from the fact that the map
$b\mapsto bb^\dagger$ is a diffeomorphism between the group $B$ and the
space of positive definite Hermitian matrices.
 \noindent \emph{Q.E.D.}
  \medskip

\noindent Notice that the element $\nu$ (4.2) characterized by Lemma 1 is
    nothing but the constant $\nu(x)$ given by Eq. (1.3).

\medskip

 \noindent {\bf Lemma 2.} For every $g\in U(n)$ and $K\in \gln$, it holds
 \be
 \Lambda(g \triangleright K)\Lambda(g \triangleright  K)^\dagger=g\Lambda(K)\Lambda(K)^\dagger g^{-1}.
 \ee

 \medskip
 \noindent {\bf Proof.}  We remark that
 \be
 \Lambda(g \triangleright K)=\Lambda_L(g \triangleright  K)
 \Lambda_R(g \triangleright K)= \Lambda_L(gK)\Lambda_R(K\Xi_R(g\Lambda_L(K))).
 \ee
Hence we have,
 \bea
 &&\Lambda(g \triangleright  K)\Lambda(g \triangleright K)^\dagger=
  \Lambda_L(gK)   \Xi_R(g\Lambda_L(K))^\dagger K^{-1}{K^{-1}}^\dagger\Xi_R(g\Lambda_L(K)) \Lambda_L(gK)^\dagger =\\
  && =g\Lambda_L(K) K^{-1}{K^{-1}}^\dagger\Lambda_L(K)^\dagger g^{-1}=
  g\Lambda_L(K)\Lambda_R(K)\Lambda_R(K)^\dagger \Lambda_L(K)^\dagger g^{-1}=
g\Lambda(K)\Lambda(K)^\dagger g^{-1}. \quad\,\,
\nonumber  \eea
 \noindent \emph{Q.E.D.}

\medskip

 \noindent {\bf Lemma 3.}  \emph{Every element $K\in \gln$ can be written as
 \be
 K=g \triangleright  (bT^{-1}),\label{rep}\ee
 where $g\in U(n)$, $b\in B$ and $T$  is in the closure of the Weyl alcove, i.e., $T\in\bar\C$.
 Moreover, for every fixed non-zero $x\in \bR$, the element $g$ can be chosen to satisfy
 the condition  $(g^{-1}v)_k\geq 0$ for all $k=1,\ldots, n$, where $v$ is the vector defined by Eq. (\ref{vx}).}

\medskip

 \noindent {\bf Proof.} For every $h\in U(n)$ and every $K\in \gln$,   the following identity
 follows from the definition (\ref{Iwasawa}) of the Iwasawa maps:
\be
\Xi_R(\Xi_R^{-1}(h\Lambda_L^{-1}(K))\Lambda_L(K))=h^{-1}.
\ee
We infer the surjectivity of the  map $\eta_K: U(n)\to U(n)$ defined for fixed $K\in \gln$ by
\be
\eta_K( g)= \Xi_R(g^{-1}\Lambda_L(K)).
\ee
Then the existence of  $g\in U(n)$, such that the matrix $T:=\Xi_R(g^{-1} \triangleright K)$
is in $\bar\C$,  is a consequence of the
relation
\be
\Xi_R (g^{-1} \triangleright K)= \Xi_R(g^{-1} \Lambda_L(K))^{-1} \Xi_R(K) \Xi_R(g^{-1} \Lambda_L(K)).
\label{N.21}\ee
Let $\tau$ be any diagonal element of $U(n)$. Then  we have
\be
g \triangleright (bT^{-1})=(g\tau) \triangleright (\tau^{-1}b\tau T^{-1}).
\ee
We thus  see that the representation (\ref{rep}) of the element $K\in\gln$
is not unique because  instead of the triple $g\in U(n),b\in B, T\in \bar\C$ we can take a triple
$g\tau\in U(n)$, $\tau^{-1}b\tau\in B$, $T\in \bar\C$ for any $\tau$. In particular, we can arrange the phases in
$\tau$ in such  a way that  the components of the vector $\tau^{-1}g^{-1}v$ become real and non-negative.
 \emph{Q.E.D.}

 \medskip

  \noindent
  {\bf Proof of Theorem 1.}  Fix  a real non-zero $x$ and parametrize $K\in\gln$  as in
  Lemma 3. By virtue of Lemma 2, the moment map constraint
  \be
  \Lambda(K)=\Lambda(g \triangleright (bT^{-1}))=\nu(x)
  \ee
  can be rewritten as
  \be
  g\Lambda(bT^{-1})\Lambda(bT^{-1})^\dagger g^{-1}=\nu(x)\nu(x)^\dagger,
  \label{rell}\ee
  or, as
  \be
  {\rm ln}(\Lambda(bT^{-1})\Lambda(bT^{-1})^\dagger)= g^{-1}{\rm ln}(\nu(x)\nu(x)^\dagger)g
  =x((g^{-1}v)  (g^{-1}v)^\dagger-\1_n).
  \ee
  Note that $\Lambda(bT^{-1})$ lies in the group $N$, for it holds
  \be
  \Lambda(bT^{-1})=\Lambda_L(bT^{-1})\Lambda_R(bT^{-1})=bTb^{-1}T^{-1}\in N.
  \ee
  Moreover, following Lemma 3, all components of the vector $g^{-1}v$ are real non-negative. Lemma 1
  then says that
  \be
   \Lambda(bT^{-1})=\nu(x);\label{con1}\ee
  \be g^{-1}v=v.
  \label{con2}\ee
  The condition (\ref{con1}) and the relation (\ref{rell}) imply  that $g\in G_x$, hence, we have so far
  proved that every solution of the moment map constraint (\ref{PLKKS})  can be written as
  $g \triangleright (bT^{-1})$ where $g\in G_x$ and $bT^{-1}$ satisfies the condition
  (\ref{con1}).

  \medskip

  \noindent In order to solve the condition (\ref{con1}), we represent $b$ as $b=\cn a$, where
  $a$ is diagonal and $\cn$ is in the group $N$. We find immediately that Eq.~(\ref{con1}) can be rewritten
  as
  \be
  \Lambda(bT^{-1})=bTb^{-1}T^{-1}=\cn T\cn^{-1}T^{-1}=\nu(x),
  \label{emm}\ee
  which means that $a$ can be arbitrary and $\cn$ has to satisfy the  condition
  \be
  \cn=\nu(x)T\cn T^{-1}.
  \label{con3}\ee
  Writing the condition (\ref{con3}) in components (with the help of Eq.~(\ref{bx})) gives
  \be
  (1- T_jT_{j+k}^{-1})\cn_{j,j+k}=\sum_{s=1}^k(1-e^{-x})e^{sx\over 2} T_{j+s}\cn_{j+s,j+k}T_{j+k}^{-1}.
  \label{j}\ee
With the replacement $j\to j-1$ and $k\to k+1$, Eq. (\ref{j}) becomes
    \bea
    (1- T_{j-1}T_{j+k}^{-1})\cn_{j-1,j+k}=\sum_{s=1}^{k+1}(1-e^{-x})e^{sx\over 2}
     T_{j-1+s}\cn_{j-1+s,j+k}T_{j+k}^{-1}=
    \nonumber\\
    = e^{x\over 2}  \sum_{s=1}^k(1-e^{-x})e^{sx\over 2}
    T_{j+s}\cn_{j+s,j+k}T_{j+k}^{-1}+(e^{x\over 2}-e^{-{x\over 2}}) T_jT_{j+k}^{-1}\cn_{j,j+k}=
    \nonumber\\
    = e^{x\over 2}(1- T_jT_{j+k}^{-1})\cn_{j,j+k}+(e^{x\over 2}-e^{-{x\over 2}}) T_jT_{j+k}^{-1}\cn_{j,j+k}=
        (e^{x\over 2} -e^{-{x\over 2}} T_jT_{j+k}^{-1})\cn_{j,j+k} .
        \label{j-1}\eea
        This equation implies that  $T_j\neq T_{j+k}$ for all $j$ and $k$, which means
  that $T$ must be a regular element of the Weyl alcove. In other words, the strict inequalities
  must hold in (\ref{strict}).
Indeed, by setting  $k=0$  we that see that $T_{j-1}\neq T_j $
(and $\cn_{j-1,j}\neq 0$) because the r.h.s. is non-zero
$(\cn_{jj}=1)$.  Then for $k=1$ we see that $T_{j-1}\neq T_{j+1}$ (and
$\cn_{j-1,j+1}\neq 0$) because the r.h.s. is non-zero, and so on.
   From Eq.~(\ref{j-1}) we can also infer that
   \be
   \cn(T)_{kl}=\prod_{m=1}^{l-k}
\frac{e^{\frac{x}{2}} T_l-e^{-\frac{x}{2}}
 T_{k+m}}{ T_l- T_{k+m-1}},
\qquad \forall k<l.
\ee
It remains to prove the last sentence of the statement of Theorem 1.   Suppose that there
exist $T_1,T_2\in \C$, $a_1,a_2\in A$ and $g \in G_x$ such that
\be
\cn(T_1)a_1T_1^{-1}= g \triangleright (\cn(T_2)a_2T_2^{-1}).
\label{rov}\ee
From Eq.~(\ref{rov}), we have obviously
\be
\Lambda_L(g \triangleright (\cn(T_2)a_2T_2^{-1}))=\Lambda_L(g\cn(T_2)a_2)
=\Lambda_L(\cn(T_1)a_1T_1^{-1})=\cn(T_1)a_1.
\label{vzt}\ee
Moreover, from  Eqs.~(\ref{N.21})  and (\ref{rov}), we obtain
\be
\Xi_R(g \triangleright (\cn(T_2)a_2T_2^{-1}))=\Xi_R(g\cn(T_2)a_2)^{-1}T_2\Xi_R(g\cn(T_2)a_2)=
 \Xi_R(\cn(T_1)a_1T_1^{-1})=T_1.
 \ee
 Since both $T_1$ and $T_2$ are in the open Weyl alcove, it follows that $\Xi_R(gn(T_2)a_2):= \Theta$
is in the maximal torus $\bT_n$ and  $T_1=T_2$ . Thus, with the help of Eq.~(\ref{vzt}), we have
 \be
 g \cn(T_2)a_2=\Lambda_L(g \cn(T_2)a_2)\Xi_R(g \cn(T_2)a_2)^{-1}= \cn(T_1)a_1\Theta^{-1}.\label{ro2}
 \ee
 Because $\Theta$ is diagonal, we have also
 \be
 \Theta^{-1}= \Xi_L( \cn(T_1)a_1\Theta^{-1})=\Xi_L(g\cn(T_2)a_2)=g,
 \ee
 hence Eq.~(\ref{ro2}) can be rewritten as
 \be
  \Theta^{-1}\cn(T_2)a_2=\cn(T_1)a_1\Theta^{-1}.
  \ee
This implies the desired statement $a_1=a_2$; and it is also worth observing that $U(n)$
factorized by its center acts
freely on the constrained manifold defined by (\ref{PLKKS}).  \emph{Q.E.D.}

 \medskip

\noindent {\bf  Remark.}  We can now  explain how we have found the Poisson-Lie analogue $\nu(x)$ (1.3)
of the Kazhdan-Kostant-Sternberg element  $\iota(x)$ (1.1).  First,    we wanted to proceed similarly to \cite{KKS},
going to a gauge where $g_R$ is diagonal.  We found that the moment map constraint (\ref{emm})
can be then satisfied only if $\nu(x)_{kk}=1$ for all $k$. Second, we wanted that the dimension
of the isotropy group $G_x$ (\ref{Gx}) be the same as in the
standard Kazhdan-Kostant-Sternberg construction.
 These two requirements lead  to the choice
of $\nu(x)$ given in (1.3).

 \section{Discussion and outlook}

  As was already mentioned in the Introduction,
 it had been known previously   that trigonometric Ruijsenaars-Schneider models
 can be   obtained   by  reduction  based    on  {\it infinite-dimensional ordinary} symmetry  \cite{GN}.
 In our approach, we have achieved essentially the same goal by means of reduction based
 on {\it finite-dimensional Poisson-Lie}  symmetry.  Apart from avoiding analytical subtleties
 of infinite-dimensional manifolds, one of the advantages of our finite-dimensional
 approach  seems to be the fact that
 the geometric picture relying on the Heisenberg double  automatically comes together  with  the integration
 algorithm for the commuting Ruijsenaars-Schneider flows.
 It appears to us that this
 cannot be obtained in a similarly  simple and direct manner in the  infinite-dimensional approach.

    \medskip

   \noindent There exists also a {\it finite-dimensional} Hamiltonian reduction treatment of the {\it complex}
trigonometric Ruijsenaars-Schneider  model utilizing a certain holomorphic,
Poisson-Lie symmetric phase space \cite{FR}.
It is not clear to us whether it is possible
 to consider a  `real form'  of that  construction  (especially  {\it before} the reduction)
 in a way which would make contact
 with our reduction. We expect, however, that the search for a  real form variant of the construction of
 \cite{FR} might be a fruitful approach to arrive at a geometric understanding of the
 {\it hyperbolic}  Ruijsenaars-Schneider model.   We plan to study this problem
 in the future.

\medskip
\noindent
Related further  papers treat rational and elliptic Ruijsenaars-Schneider models as well,
applying various versions of classical
 Hamiltonian reduction, but none of the works published to date  use compact,
 finite-dimensional Poisson-Lie groups to obtain directly the real
trigonometric model, which we achieved here.
The reader may consult \cite{AFM1,AFM2,Nekr,Fock+} and the references therein.

 \medskip

  \noindent Since
   our principal guideline in writing this paper was   an effort to address also  the readers
   who  are not experts in  Poisson-Lie geometry, we have not  worked out here  several further
    issues  that would require a deeper  preliminary exposition  of  the general theory of  Poisson-Lie symmetry.
    To fill this gap, we are  preparing a continuation of this article. We  shall  present there  a geometric
  interpretation of the Ruijsenaars duality \cite{SR-CMP,SR-RIMS,SR-CRM} that
links together two different real forms of the complex trigonometric Ruijsenaars-Schneider model.
We shall treat the duality in terms of two  gauge slices  (of group theoretic origin) in the Poisson-Lie reduction,
generalizing  the picture put forward in \cite{Nekr,Fock+} to understand
the duality  between the  trigonometric  Sutherland model
   and a certain real form of the complex rational Ruijsenaars-Schneider model \cite{SR-RIMS}.
   As a byproduct, we shall directly obtain
    the  dual Lax matrix and the dual commuting flows as well.  Finally, we shall give
    an account of the non-relativistic limit
   of the trigonometric Ruijsenaars-Schneider  model  in the geometric perspective.
    Thereby we shall explain how the Poisson-Lie
   reduction described in this paper can be viewed not only as a conceptual generalization
   of  the standard Kazhdan-Kostant-Sternberg reduction but also as  its natural one-parameter deformation.

 \medskip

  \noindent
  We are aware of the fact that there exists a large literature also on the
  quantum group aspects of the quantum mechanical trigonometric Ruijsenaars-Schneider model,
  but to discuss this is outside the scope of the present work. See, for example,
  \cite{SR-CRM,EK,Noum} and
  references therein.
  The relationship to the quantization of our Poisson-Lie reduction
  may be worth further investigation.

\bigskip
\noindent{\bf Acknowledgements.}
L.F. was supported in part
by the Hungarian
Scientific Research Fund (OTKA grant
 T049495) and by the EU network `ENIGMA'
(contract number MRTN-CT-2004-5652).
He is also grateful to the IML for an invitation,
and thanks J. Balog for discussions.


\begin{thebibliography}{9}

\bibitem{AM}
Alekseev, A.Yu.,  Malkin, A. Z.:
Symplectic structures associated to Lie-Poisson groups.
Commun. Math. Phys. {\bf 162},  147-174 (1994)

\bibitem{AFM1}
Arutyunov, G.E., Frolov,  S.A., Medvedev,  P.B.:
Elliptic Ruijsenaars-Schneider model via the Poisson reduction of
the affine Heisenberg double.
J. Phys. A: Math. Gen. {\bf 30},   5051-5063 (1997)

\bibitem{AFM2}
Arutyunov, G.E., Frolov,  S.A., Medvedev,  P.B.:
Elliptic Ruijsenaars-Schneider model from the cotangent bundle over the two-dimensional current group.
J. Math. Phys. {\bf 38}, 5682-5689 (1997)

\bibitem{Cal}
Calogero, F.: Solution of the one-dimensional $N$-body problem
with quadratic and/or inversely quadratic pair potentials.
J. Math. Phys. {\bf 12},  419-436 (1971)

\bibitem{EK}
Etingof, P.I., Kirillov Jr., A.A.:
Macdonald's polynomials and representations of quantum groups.
Math. Res. Lett. {\bf 1}, 279-294 (1994)

\bibitem{Fock+}
Fock, V.,  Gorsky, A., Nekrasov,  N.,  Rubtsov,  V.:
Duality in integrable systems and gauge theories.
JHEP {\bf 0007},  028 (2000)

\bibitem{FR}
Fock, V.V.,   Rosly, A.A.:
Poisson structure on moduli of flat connections on Riemann surfaces and the $r$-matrix.
In: Moscow Seminar in Mathematical Physics,
AMS Transl. Ser. 2, Vol.~191, 1999, pp. 67-86


\bibitem{GN}
Gorsky, A.,  Nekrasov,  N.:
Relativistic Calogero-Moser model as gauged WZW theory.
Nucl. Phys. {\bf B436}, 582-608  (1995)

\bibitem{KKS}
Kazhdan, D.,  Kostant,  B.,  Sternberg, S.:
Hamiltonian group actions and dynamical systems of Calogero type.
Comm. Pure Appl. Math. {\bf XXXI}, 481-507 (1978)

\bibitem{K}
Klim\v c\'\i k, C.:
On moment maps associated to a twisted Heisenberg double.
 Rev. Math. Phys. {\bf 18},  781-821 (2006)

\bibitem{Lu-red}
 Lu, J.-H.:
Moment maps and reduction of Poisson actions.
In: Proc. of
Sem. Sud-Rodanian de Geometrie \`a Berkeley (1989),
Springer-Verlag MSRI Publ., Vol.~20,  1991, pp. 209-226


\bibitem{Mos}
 Moser, J.: Three integrable Hamiltonian systems connected with
isospectral deformations. Adv. Math. {\bf 16},  197-220 (1975)

\bibitem{Nekr}
Nekrasov, N.:
Infinite-dimensional algebras, many-body systems and gauge theories.
In: Moscow Seminar in Mathematical Physics, AMS Transl. Ser. 2,
Vol.~191, 1999, pp. 263-299

\bibitem{Noum}
Noumi, M.:
Macdonald's symmetric polynomials  as zonal spherical functions on some quantum homogeneous spaces.
Adv. Math. {\bf 123}, 16-77 (1996)

\bibitem{OP1}
Olshanetsky, M.A.,   Perelomov, A.M.:
Explicit solution of the Calogero model in the classical case and
geodesic flows on symmetric spaces of zero curvature.
Lett. Nouvo Cim. {\bf 16},  333-339 (1976)

\bibitem{OP2}
Olshanetsky, M.A.,   Perelomov, A.M.:
Explicit solutions of some completely integrable systems.
Lett. Nouvo Cim. {\bf 17},  97-101 (1976)

\bibitem{ST}
Semenov-Tian-Shansky, M.A.:  Dressing transformations and Poisson groups actions.
Publ. RIMS {\bf 21}, 1237-1260  (1985)

\bibitem{Sut}
Sutherland, B.: Exact results for a quantum many body problem in
one dimension. Phys. Rev. {\bf A4}, 2019-2021  (1971)


\bibitem{SR-CMP}
Ruijsenaars, S.N.M.:
Action-angle maps and scattering theory for some finite-dimensional
integrable systems I. The pure soliton case.
Commun. Math. Phys. {\bf 115},  127-165 (1988)

\bibitem{SR-RIMS}
Ruijsenaars, S.N.M.:
Action-angle maps and scattering theory for some finite-dimensional
integrable systems III. Sutherland type systems and their duals.
Publ. RIMS  {\bf 31}, 247-353 (1995)

\bibitem{SR-CRM}
Ruijsenaars, S.N.M.:
Systems of Calogero-Moser type.
In: Proceedings of the 1994 CRM--Banff Summer School `Particles and Fields',
Springer, 1999,
pp.~251-352

\bibitem{RS}
 Ruijsenaars, S.N.M.,  Schneider, H.:
A new class of integrable models and their relation to solitons.
Ann. Phys. (N.Y.) {\bf 170}, 370-405  (1986)


\bibitem{DV}
van Diejen, J.F.,  Vinet, L.:
The quantum dynamics of the compactified trigonometric Ruijsenaars-Schneider model.
Commun. Math. Phys. {\bf 197},  33-74 (1998)


\bibitem{Zakr}  Zakrzewski, S.:
Free motion on the Poisson $SU(N)$ group.
J. Phys. A: Math. Gen. {\bf 30},  6535-6543 (1997)

\end{thebibliography}
\end{document}